\documentclass[twocolumn,twoside,10pt,aps,prl,superscriptaddress,amsmath,amssymb,longbibliography]{revtex4-2}
\usepackage[switch]{lineno}
\usepackage[english]{babel}
\usepackage{mathrsfs}
\usepackage{graphicx}
\usepackage{amsmath}
\usepackage{amssymb}
\usepackage{amsfonts}
\usepackage{color}
\usepackage{leftidx}
\usepackage{verbatim}

\def\bra#1{\left\langle{#1}\right|}
\def\ket#1{\left|{#1}\right\rangle}
\def\braket#1#2{\left\langle{{#1}}\mathrel{\left|{\vphantom{{#1}{#2}}}\right.\kern-\nulldelimiterspace}{{#2}}\right\rangle}

\DeclareUnicodeCharacter{2062}{} 

\begin{document}

\title{Ramsey Spectroscopy via Symmetry-Protected Destructive Many-Body Interferometry}

\author{Sijie Chen}
\affiliation{Laboratory of Quantum Engineering and Quantum Metrology, School of Physics and Astronomy, Sun Yat-Sen University (Zhuhai Campus), Zhuhai 519082, China}
\affiliation{Institute of Quantum Precision Measurement, State Key Laboratory of Radio Frequency Heterogeneous Integration, College of Physics and Optoelectronic Engineering, Shenzhen University, Shenzhen 518060, China}
\author{Jiahao Huang}
\affiliation{Laboratory of Quantum Engineering and Quantum Metrology, School of Physics and Astronomy, Sun Yat-Sen University (Zhuhai Campus), Zhuhai 519082, China}
\affiliation{Institute of Quantum Precision Measurement, State Key Laboratory of Radio Frequency Heterogeneous Integration, College of Physics and Optoelectronic Engineering, Shenzhen University, Shenzhen 518060, China}
\author{Min Zhuang}
\altaffiliation{Email: mmzhuang@szu.edu.cn}
\affiliation{Institute of Quantum Precision Measurement, State Key Laboratory of Radio Frequency Heterogeneous Integration, College of Physics and Optoelectronic Engineering, Shenzhen University, Shenzhen 518060, China}
\author{Chaohong Lee}
\altaffiliation{Email: chleecn@szu.edu.cn, chleecn@gmail.com}
\affiliation{Institute of Quantum Precision Measurement, State Key Laboratory of Radio Frequency Heterogeneous Integration, College of Physics and Optoelectronic Engineering, Shenzhen University, Shenzhen 518060, China}
\affiliation{Quantum Science Center of Guangdong-Hong Kong-Macao Greater Bay Area (Guangdong), Shenzhen 518045, China}
\begin{abstract}
Ramsey spectroscopy, a fundamental tool in both basic science and practical applications, is inevitably subject to several detrimental effects.
Here we propose a symmetry-protected destructive many-body interferometry (SPDMBI) for Ramsey spectroscopy, which successfully mitigates the spectral shift caused by interparticle interaction, noise, decoherence and experimental imperfection.
Through matching the symmetry of the input states and the Hamiltonian, the SPDMBI-based Ramsey spectroscopy yields an antisymmetric spectrum, whose antisymmetric point exactly determines the resonance frequency.
In such a Ramsey spectroscopy, the population difference under resonance is always zero, which is a result of destructive quantum interferometry.
To demonstrate its versatility, we showcase successful applications of symmetry-protected Ramsey spectroscopy in measuring both time-independent and time-dependent signals.
Our protocol can improve the performance of Ramsey spectroscopy, which offers a pathway for various high-precision quantum sensors.
\end{abstract}

\date{\today}
\maketitle
%
{\it Introduction. --}
%
Ramsey spectroscopy, a cornerstone technique in precision measurement, is crucial for both fundamental science and practical technology~\cite{NFRPR1950,NFRRMP1990,CCT1992,CLDRMP2017,SS2022,LPRMP2018,huang2024entanglementenhanced}. 
By precisely measuring the transition frequencies between energy levels, Ramsey spectroscopy enables in-depth studies of the structure and properties of matter, leading to a wide range of applications in quantum technologies~\cite{DDNP2024,SDBNRP2024,JDRMP2024}, astrophysics~\cite{KLRMP2003,EARMP2010,PRNA2019,MLNC2024}, optical spectroscopy~\cite{NBRMP1982,KEDRMP2016}, and other fields~\cite{RTNC2020,WKRMP2023}.
These capabilities are essential for investigating fundamental physics~\cite{MZQ2022} and have practical implications in fields such as geophysics~\cite{SB2022}, materials science~\cite{KBNP2017}, and biomedical sensing~\cite{AN2023}.

In practice, noise, decoherence and experimental imperfection are inevitable and always exert detrimental effects on Ramsey spectroscopy. 
Dynamical decoupling~\cite{CLDRMP2017,DSRMP2016,MORMP2025,HYCPR1954,SMRSI1958,LVPRL821999,LVPRL831999,GDLScience2010,LJPRA2011,NEPRAppl2023,SCCPB2025}, typically achieved through carefully engineered sequences of control pulses to mitigate noise, has become essential for quantum sensing.
However, most approaches focus on single-particle systems or non-interacting systems. 
In Ramsey spectroscopy involving interacting particles, while increasing interparticle interactions can theoretically improve measurement precision, it often compromises accuracy~\cite{YBBPRA2006,HZPRX2020,JHarxiv2022}.
The nonlinear nature of these interactions tends to introduce additional spectral shifts, which can be exacerbated by various noise sources, including white noise, Rabi frequency fluctuations, timing inaccuracies, readout noise, and decoherence. 
These factors can easily overwhelm spectroscopic signals. 
Consequently, designing distinct dynamical decoupling sequences for each noise source becomes necessary. 
This leads to a critical question: can we develop a universal Ramsey spectroscopy protocol that remains robust against control inaccuracies and decoherence in interacting multiparticle quantum systems?

In this Letter, we propose a symmetry-protected destructive many-body quantum interferometry (SPDMBI) to address the above challenge.
Drawing on the principles of symmetry protection in quantum adiabatic evolution~\cite{MZAP2020,THNJP2022}, topological phases~\cite{CKCRMP2016} 
and subspaces~\cite{CRPRR2023}, our SPDMBI protocol is specifically designed to match the system Hamiltonian's symmetry with symmetric input states. 
This approach can mitigate spectral shifts caused by interparticle interactions, several decoherence channels, and experimental imperfections, thereby maintaining the measurement accuracy.
The protocol's versatility enables direct application across various sensing modalities, including Rabi~\cite{CCT2019} and Ramsey spectroscopy for time-independent signals~\cite{NFRPR1950,NFRRMP1990,SCL2010}, as well as lock-in amplifiers~\cite{SKNature2011,MZPRX2021,MZQF2024,SCCP2024} and heterodyne measurements~\cite{SSScience2017,JMBScience2017,JMNC2021,NSPRL2023} for time-dependent signals. 

Moreover, when inputting symmetric entangled states, our SPDMBI protocol can further improve measurement precision without compromising accuracy, as detailed in the companion paper~\cite{SCPRA2025}.
Using currently available techniques, our protocol can be readily implemented in various quantum systems, including Bose-condensed atoms~\cite{ADCRMP2009,MWMRMP2020}, trapped ions~\cite{DLRMP2003,CMRMP2021}, nitrogen-vacancy centers in diamond~\cite{JFBRMP2020,JDRMP2024}, doped spins in semiconductors~\cite{GBRMP2023}, and even ``artificial atoms'' like superconducting qubits~\cite{ABRMP2021,JBNP2011,YSPRAppl2019,NEPRAppl2023}.
%
\begin{figure}[!htp]
\includegraphics[width=\columnwidth]{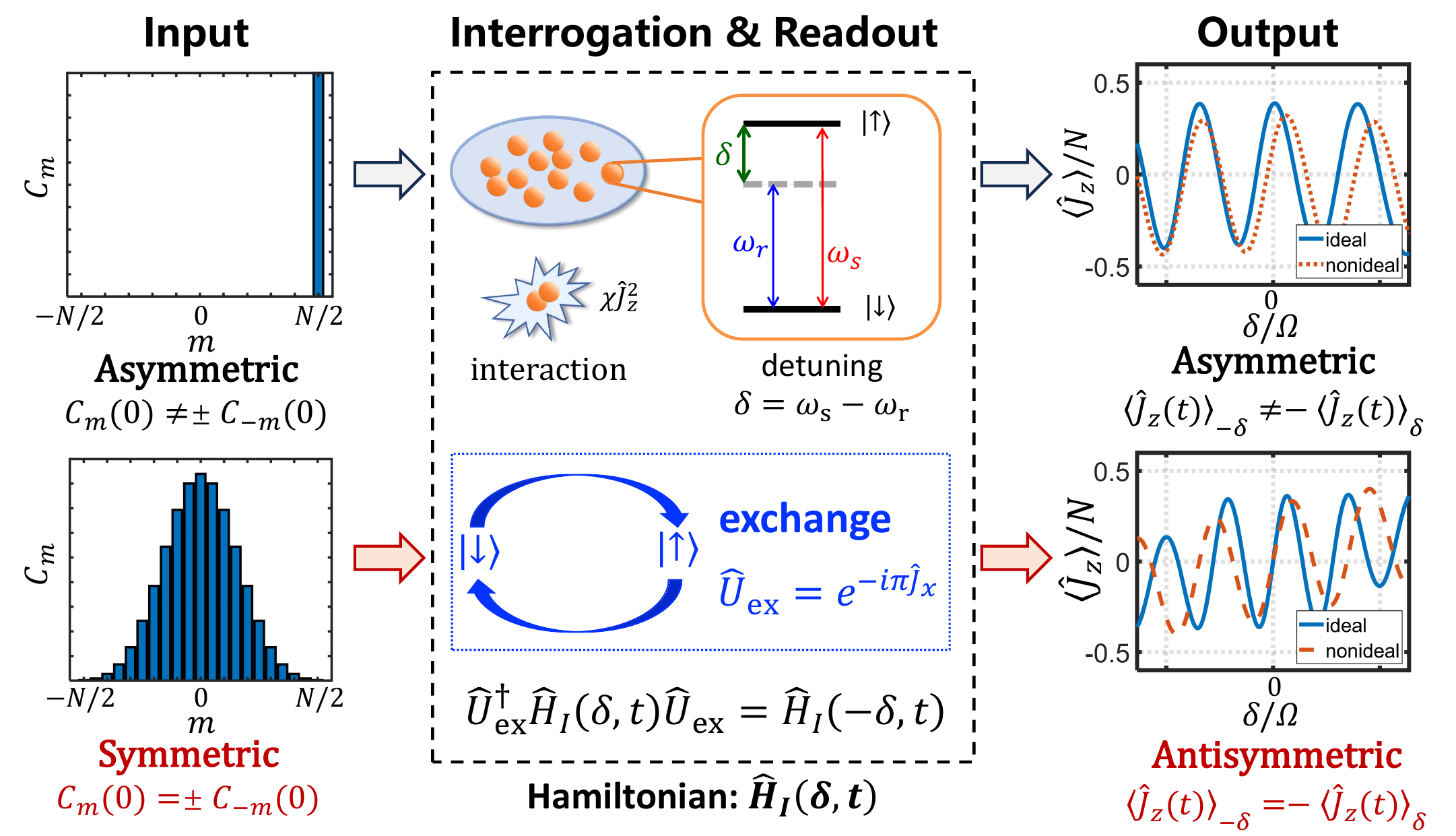}
\caption{\label{Fig1}(color online).
Schematic diagram for symmetry-protected destructive many-body interferometry. 
In the interaction picture, the systems obeys a Hamiltonian $\hat{H}_\textrm{I}(\delta,t)$ of symmetry~\eqref{H_sym}. 
Inputting symmetric states satisfying Eq.~\eqref{Coef}, the output spectrum is always antisymmetric with respect to $\delta$ (solid line). 
While inputting other states, the output spectrum becomes asymmetric.
}
\end{figure}

{\it Symmetry-protected destructive many-body interferometry (SPDMBI). --}
We consider an ensemble of $N$ two-mode bosonic particles, where each two-mode particle is labeled as a half-spin with $\ket{\uparrow}$ and $\ket{\downarrow}$.
Using the Schwinger representation, the system can be characterized by a collective spin $\hat{\vec{J}}=\left(\hat{J}_x,\hat{J}_y,\hat{J}_z\right)^{T}$ with $\hat{J}_x =\frac{1}{2}
\left(\hat{a}^{\dagger}\hat{b}+\hat{a}\hat{b}^{\dagger}\right)$, $
\hat{J}_y=\frac{i}{2}\left(\hat{a}^{\dagger}\hat{b}-\hat{a}\hat{b}^{\dagger}\right)$, $\hat{J}_z=\frac{1}{2}\left(\hat{a}^{\dagger}\hat{a}-\hat{b}^{\dagger}\hat{b}\right)$,
where $\hat{a}$ and $\hat{b}$ denote the annihilation operators for $\ket{\uparrow}$ and $\ket{\downarrow}$, respectively.
The system states can be represented in terms of the Dike basis $\ket{J,m}$, where $J=N/2$ and $m=\{-J,-J+1,...,J-1,J\}$.
In general, the total Hamiltonian for precision spectroscopy can be written as $\hat{H}=\hat{H}_\textrm{a}+\hat{H}_\textrm{s}+\hat{H}_\textrm{r}$.
The first term $\hat{H}_\textrm{a}=\chi\hat{J}_z^2$ describes a non-linear interaction with $\chi$ characterizing the strength~\cite{JHarxiv2022,TParXiv2025}.
The second term $\hat{H}_\textrm{s}=\vec{S}(\omega_s,t)\cdot\hat{\vec{J}}$ denotes the coupling between the probe and the signal field to be estimated.
The third term $\hat{H}_\textrm{r}=\vec{R}(\omega_r,t)\cdot\hat{\vec{J}}$ represents the auxiliary control with additional fields~\cite{SKNature2011,MZPRX2021,JHarxiv2022,SCCP2024}.
Here, $\vec{S}(\omega_s,t)$ and $\vec{R}(\omega_r,t)$ are three-dimensional vectors oscillating versus time $t$ with oscillation frequencies $\omega_s$ and $\omega_r$.
In the interaction picture, the system obeys a Hamiltonian in forms of
$\hat{H}_\textrm{I}(\delta, t)=f_1\hat{J}_x+f_2\hat{J}_y+f_3\hat{J}_z+e_1\hat{J}_y\hat{J}_z+e_2\hat{J}_z\hat{J}_x+e_3\hat{J}_x\hat{J}_y+g_1\hat{J}_x^2+g_2\hat{J}_y^2+g_3\hat{J}_z^2$~\cite{SKNature2011,MZPRX2021,NFRPR1950,SSScience2017,JMBScience2017,TZWRPP2018},
where $x_i\equiv x_i(\delta,t)$ (for $x=e,f,g$ and $i=1,2,3$) with $\delta=\omega_s-\omega_r$ the detuning between the target and reference signals.
Therefore, given $\omega_r$, one estimates $\delta$ to infer $\omega_s$.
%
For an initial state $\ket{\Psi(0)}=\sum_{m=-J}^{J}C_m(0)\ket{J,m}$, the instantaneous state $\ket{\Psi(\delta,t)}=\sum_{m=-J}^{J}C_m(\delta,t)\ket{J,m}$ obeys the Schr\"{o}dinger equation $i\frac{\partial\ket{\Psi(\delta,t)}}{\partial t}=\hat{H}_I(\delta,t){\ket{\Psi(\delta,t)}}$ (in units of $\hbar=1$).
The half-population difference can be obtained via $\langle \hat J_z(t)\rangle_{\delta}=\textrm{Tr}[\hat \rho(\delta,t) \hat J_z ]$ with $\hat \rho(\delta,t)=\ket{\Psi(\delta,t)}\bra{\Psi(\delta,t)}$ and $\textrm{Tr}[\hat{X}]$ denoting the trace of $\hat{X}$.

In most quantum sensing scenarios, such as Rabi and Ramsey spectroscopy, the Hamiltonian $\hat{H}_\textrm{I}(\delta,t)$ possesses intriguing symmetry [see Fig.~\ref{Fig1}].
Under the transformation $\hat{a}\leftrightarrow\hat{b}$, which is described by $\hat{U}_\textrm{ex}=e^{-i\pi\hat{J}_x}$, we have $\hat{U}_\textrm{ex}^{\dagger}\hat{J}_x\hat{U}_\textrm{ex}=\hat{J}_x$ and $\hat{U}_\textrm{ex}^{\dagger}\hat{J}_{y,z}\hat{U}_\textrm{ex}=-\hat{J}_{y,z}$. 
In both Rabi and Ramsey spectroscopy~\cite{MZPRX2021,JHarxiv2022}, the coefficients always satisfy $f_1=f_1^\textrm{even}$, $e_1=e_1^\textrm{even}$, $g_{1,2,3}=g_{1,2,3}^\textrm{even}$, $f_{2,3}=f_{2,3}^\textrm{odd}$ and $e_{2,3}=e_{2,3}^\textrm{odd}$, the Hamiltonian becomes
\begin{eqnarray}\label{HwI}
\hat{H}_\textrm{I}(\delta,t)&=&f_1^\textrm{even}\hat{J}_x+f_2^\textrm{odd}\hat{J}_y+f_3^\textrm{odd}\hat{J}_z\nonumber\\
& &+e_1^\textrm{even}\hat{J}_y\hat{J}_z+e_2^\textrm{odd}\hat{J}_z\hat{J}_x+e_3^\textrm{odd}\hat{J}_x\hat{J}_y\nonumber\\
& &+g_1^\textrm{even}\hat{J}_x^2+g_2^\textrm{even}\hat{J}_y^2+g_3^\textrm{even}\hat{J}_z^2,
\end{eqnarray}
and it has the symmetry
\begin{equation}\label{H_sym} \hat{U}_\textrm{ex}^{\dagger}\hat{H}_I(\delta,t)\hat{U}_\textrm{ex}=\hat{H}_I(-\delta,t).
\end{equation}
Here $x_i^\textrm{even}(-\delta,t)=x_i^\textrm{even}(\delta,t)$ and $x_i^\textrm{odd}(-\delta,t)=-x_i^\textrm{odd}(\delta,t)$ are symmetric and antisymmetric functions versus $\delta$.
Under the symmetry~\eqref{H_sym}, one can obtain robust antisymmetric spectra using symmetric input states. 
%
For a symmetric initial state, $\ket{\Psi}_S=\sum_{m=-J}^{J}C_m(0)\ket{J,m}$ with the coefficients
\begin{equation}\label{Coef}
    C_m(0)=\pm C_{-m}(0),
\end{equation}
we have $\hat{U}_\textrm{ex}^{\dagger} \hat{\rho}(0)\hat{U}_\textrm{ex}=\hat{\rho}(0)$ with $\hat{\rho}(0)=\ket{\Psi_0}\bra{\Psi_0}$~\cite{SCPRA2025}.
Actually, one can easily find that these symmetric states are eigenstates of $\hat{U}_\textrm{ex}$ and $\hat{J}_x$.
Therefore, the half-population difference is antisymmetric about the detuning $\delta$ at any time $t$, i.e., 
\begin{equation}\label{antisym}
    \langle \hat J_z(t)\rangle_{\delta}=-\langle \hat J_z(t)\rangle_{-\delta}.
\end{equation}
Particularly, at the resonance point $\delta=0$, the half-population difference always vanishes, i.e. $\langle \hat J_z(t)\rangle_{\delta=0}=0$.
Thus we call the above destructive interferometry as symmetry-protected destructive many-body interferometry (SPDMBI).

Using the symmetry of the initial state $\hat{\rho}(0)$ and the Hamiltonian $\hat{H}_\textrm{I}(\delta,t)$, the relation~\eqref{antisym} can be analytically derived. 
Given a symmetric initial state $\hat{\rho}(0)=\hat{U}_\textrm{ex}^\dagger\hat{\rho}(0)\hat{U}_\textrm{ex}$, according to the Schr\"{o}dinger equation and the Hamiltonian symmetry~\eqref{H_sym}, we have $\hat{U}_\textrm{ex}^\dagger\partial_t\hat{\rho}(\delta,t)\hat{U}_\textrm{ex}=\partial_t\hat{\rho}(-\delta,t)$ and $\hat{U}_\textrm{ex}^\dagger\hat{\rho}(-\delta,t)\hat{U}_\textrm{ex}=\hat{\rho}(\delta,t)$.
Therefore, we have $\langle\hat{J}_z(t)\rangle_{-\delta}=\textrm{Tr}[\hat{\rho}(-\delta,t)\hat{J}_z]=\textrm{Tr}[\hat{U}_\textrm{ex}^\dagger\hat{\rho}(\delta,t)\hat{U}_\textrm{ex}\hat{J}_z]=\textrm{Tr}[\hat{U}_\textrm{ex}\hat{J}_z\hat{U}_\textrm{ex}^\dagger\hat{\rho}(\delta,t)]=-\textrm{Tr}[\hat{J}_z\hat{\rho}(\delta,t)]=-\langle\hat{J}_z(t)\rangle_{\delta}$, which gives a proof for the relation~\eqref{antisym}.
This result is valid for arbitrary Hamiltonian including time-dependent Hamiltonian in forms of Eq.~\eqref{HwI}.
Moreover, Eq.~\eqref{antisym} also holds for typical noisy systems. 
Under Markovian noises, the system can be described by a Lindblad master equation~\cite{JHSR2016}
\begin{equation}\label{Lindblad}
    \frac{\partial\hat{\rho}(\delta,t)}{\partial t}=-i\left[\hat{H}_I(\delta,t),\hat{\rho}(\delta,t)\right]+\mathcal{L}\hat{\rho}(\delta,t),
\end{equation}
%
where $\mathcal{L}\hat{\rho}=\sum_{k}\gamma_k\left(\hat{\mathcal{L}}_k\hat{\rho}\hat{\mathcal{L}}_k^\dagger-\frac{1}{2}\left\{\hat{\mathcal{L}}_k^\dagger\hat{\mathcal{L}}_k,\hat{\rho}\right\}\right)$ describes the decoherence.
If the Lindblad operator satisfies $\hat{U}_\textrm{ex}^{\dagger}\mathcal{L}\hat{\rho}(0)\hat{U}_\textrm{ex}=\mathcal{L}\hat{\rho}(0)$ (such as collective dephasing~\cite{GFPRA2010, KPPRA2013} and balanced atom losses~\cite{JLNC2019,JHSR2016}), antisymmetric spectra satisfying Eq.~\eqref{antisym} still appear under noisy environments.

\begin{figure}[!htp]
\includegraphics[width=1\columnwidth]{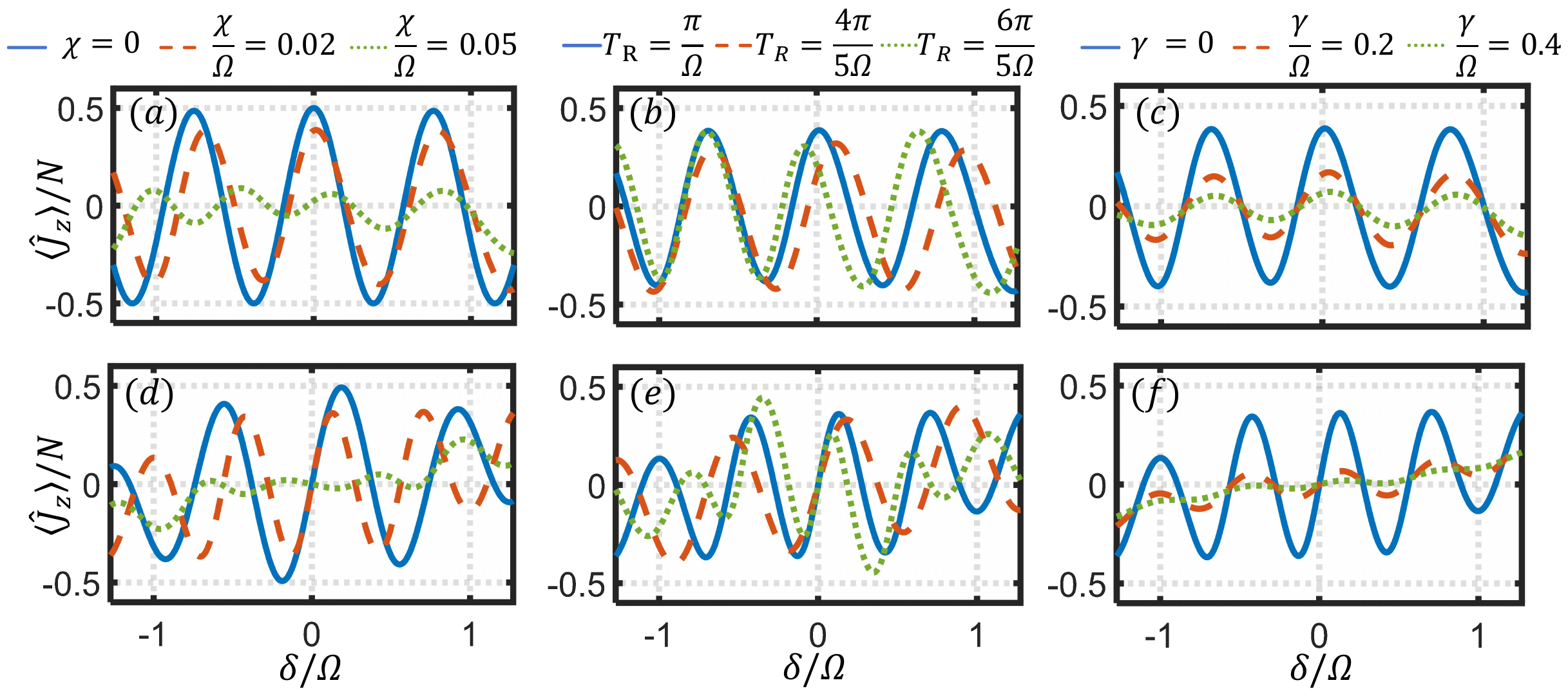}
\caption{\label{Fig2}(color online).
Many-body Ramsey spectroscopy.
Top row: inputting an asymmetric state $\ket{\Psi(0)}=\ket{J,J}$ under different (a) interaction strength $\chi$, (b) pulse duration $T_R/2$, and (c) correlated dephasing rate $\gamma$.
Bottom row: the SPDMBI with a symmetric state $\ket{\Psi(0)}=[({\ket{\uparrow}+\ket{\downarrow}})/\sqrt2]^{\otimes N}$
under different (d) interaction strength $\chi$, (f) readout time $\tau_\textrm{ro}$, and (g) correlated dephasing rate $\gamma$.
For (a) (d), $T_R=\frac{\pi}{\Omega}$, $\gamma=0$. 
For (b) (e), $\chi=0.02\Omega$, $\gamma=0$. 
For (c) (f), $T_R=\frac{\pi}{\Omega}$, $\chi=0.02\Omega$.
Here, we choose the particle number $N=20$ and the free evolution time $T_f=2\pi/\Omega$.
}
\end{figure}

If the initial state is not a symmetric state (i.e. $C_m(0) \neq \pm C_{-m}(0)$), the inter-particle interaction and experimental imperfections result in an asymmetric spectrum. 
While for the symmetric initial state~\eqref{Coef}, the spectrum is always antisymmetric and thus the resonance point can be accurately identified.
Our SPDMBI can be applied to a wide range of scenarios with strongly interacting many-body systems~\cite{LPRMP2018, huang2024entanglementenhanced, PhysRevA.108.062611} and it is robust against realistic experimental imperfections, such as, the Rabi frequency fluctuation, inaccurate time control and decoherence. 
%
%
Below we will show some typical applications of our SPDMBI.
More applications of our SPDMBI can be found in the companion paper~\cite{SCPRA2025}.

{\it Applications in detecting time-independent signals. --}
We first discuss applications of SPDMBI-based Ramsey spectroscopy in measuring time-independent signals.
We consider an ensemble of $N$ spin-$1/2$ interacting particles in a static magnetic field $B_z\hat{z}$ to be measured and a control magnetic field $B_\perp(t)[\cos(\omega_r t)\hat{x}+\sin(\omega_r t)\hat{y}]$.
Based upon Rabi spectroscopy, Ramsey spectroscopy changes the interaction zone into a free evolution sandwiched by two short $\frac{\pi}{2}$-pulse interaction zones~\cite{NFRPR1950}. 
Therefore, the Hamiltonian can be written as 
$\hat{H}_{R_0}=\omega_s\hat{J}_z+\Omega(t)[\cos(\omega_r t)\hat{J}_x+\sin(\omega_r t)\hat{J}_y]$, where $\omega_s=\gamma_g B_{z}$, the Rabi frequency \begin{eqnarray}\label{OmeRam}
\Omega(t)=\left\{
\begin{array}{rl}
\Omega,&0\leq t<T_\Omega,\\
0,&T_\Omega\leq t<T_f+T_\Omega,\\
\Omega,&T_f+T_\Omega\leq t\leq T_f+2T_\Omega,
\end{array}
\right.
\end{eqnarray}
with the free evolution time $T_f$, the duration of a $\pi/2$ pulse $T_\Omega=\pi/(2\Omega)$ and $\Omega=\gamma_g B_\perp$. 
Transforming $\hat{H}_{R_0}$ into the interaction picture with respect to $\hat{H}_0=\omega_r\hat{J}_z$ and compensating the linear shift, the system obeys~\cite{JHarxiv2022}
\begin{equation}\label{Rabi1}
\hat{H}_\textrm{R}=\chi\hat{J}_z^2+\delta\hat{J}_z+\Omega(t)\hat{J}_x
\end{equation}
with the detuning $\delta=\omega_s-\omega_r$ and the interaction strength $\chi$.
In conventional Ramsey spectroscopy, the initial state is chosen as $\ket{\Psi(0)}=\left(\ket{\downarrow}\right)^{\otimes N}=\ket{J,-J}$ and evolves to an equal superposition state $\left((\ket{\uparrow}+i\ket{\downarrow})/{\sqrt 2}\right)^{\otimes N}$ after the first $\pi/2$ pulse. However, this state does not satisfy Eq.~\eqref{Coef}.
After the free evolution with $T_f$ and the second $\pi/2$ pulse, $\langle\hat{J}_z\rangle$ is finally detected for different $\omega_r$~\cite{KAFQST2018,JHarxiv2022}.
In the absence of interaction ($\chi=0$), the Ramsey spectrum is symmetric about the detuning $\delta=\omega_s-\omega_r$, see Fig.~\ref{Fig2}~(a).
While in the case of nonzero $\chi$, the resonance peak shifts with the increase of $\chi$, which brings a systematic error in determining the resonance frequency~\cite{YBBPRA2006}.
%
In addition, inaccurate time control [Fig.~\ref{Fig2}~(b)] and dephasing [Fig.~\ref{Fig2}~(c)] will amplify the shift.
In contrast, we find that the SPDMBI can improve accuracy.
By entering a symmetric state $\ket{\Psi(0)}=\ket{\Psi}_S=\left((\ket{\uparrow}+\ket{\downarrow})/{\sqrt 2}\right)^{\otimes N}$, the Ramsey spectrum becomes antisymmetric and the antisymmetric point does not have any shift induced by the interaction $\chi$, see Fig.~\ref{Fig2}~(d).
Promisingly, the SPDMBI is robust against inaccurate time control of $T_R$ [Fig.~\ref{Fig2}~(e)] and dephasing [Fig.~\ref{Fig2}~(f)].
%
%
%

{\it Applications in detecting time-dependent signals. --}
The precise measurement of time-dependent signals within noisy background is of great significance.
The quantum lock-in amplifiers have been successfully demonstrated and widely employed to measure time-dependent alternating signals~\cite{SKNature2011,MZPRX2021,SCCP2024,MZQF2024}.
Here we consider a many-body lock-in amplifier, in which the coupling between the probe and the external field can be described by ${\hat{H}}_\textrm{int}(t)=M(t) \hat{J}_{z}$, where $M(t)=\emph{S}(t)+\emph{N}_{o}(t)$ consists of the target signal $\emph{S}(t)=\gamma_g B_\textrm{ac}\sin(\omega_s t)$ and the noise $\emph{N}_{o}(t)$.
To implement quantum lock-in measurement, one can apply a reference signal $\hat{H}_\textrm{ref}=\Omega_\pi(t)\hat{J}_x$ for mixing the target and reference signals~\cite{MZPRX2021}. 
Usually, $\hat{H}_\textrm{ref}$ can be chosen as a sequence of periodic $\pi$ pulses that does not commute with ${\hat{H}}_\textrm{int}(t)$~\cite{SKNature2011,MZPRX2021,SCCP2024}.
%
In experiments, the $\pi$-pulse sequences are usually chosen as square waves with equidistant spacing $\tau_r$ and pulse length $T_{\Omega}$.
That is, the Rabi frequency is set as
$\Omega_\pi(t)=\pi/T_\Omega$ for $|t-(l-\lambda)\tau_r|\leq T_{\Omega}/2$, otherwise $\Omega_\pi(t)=0$,
%
where $l=1,2,\cdots,L$ ($L$ denoting the pulse number), and $(1-\lambda)\tau_r$ is the time when the first $\pi$-pulse is applied.
And after $L\tau_r$ evolution time, a $\pi/2$ pulse along $x$ direction is imposed for readout, which also satisfies our SPDMBI.
For the periodic dynamical decoupling (PDD) and Carr-Purcell (CP) sequences, we have $\lambda=0$ and $\lambda=1/2$~\cite{SKNature2011,CLDRMP2017,MZPRX2021,SCCP2024}, respectively.
%

In the interaction picture with respect to $\Omega_\pi(t)\hat{J}_{x}$, the original Hamiltonian $\hat{H}_\textrm{LA}^0(t)=\chi\hat{J}_z^2+M(t)\hat{J}_z+\Omega_\pi(t)\hat J_x$ transforms into
\begin{equation}\label{HIlock0}
 \hat{H}_\textrm{LA}(t)=\chi\hat{J}_\alpha^2(t)+M(t)\hat{J}_\alpha(t),
\end{equation}
where $\hat{J}_\alpha(t)=\cos[\alpha(t)]\hat{J}_z+\sin[\alpha(t)]\hat{J}_y$ and $\alpha(t)=\int_0^t\Omega_{\pi}(t')dt'$.

\begin{figure}[!htp]
\includegraphics[width=1\columnwidth]{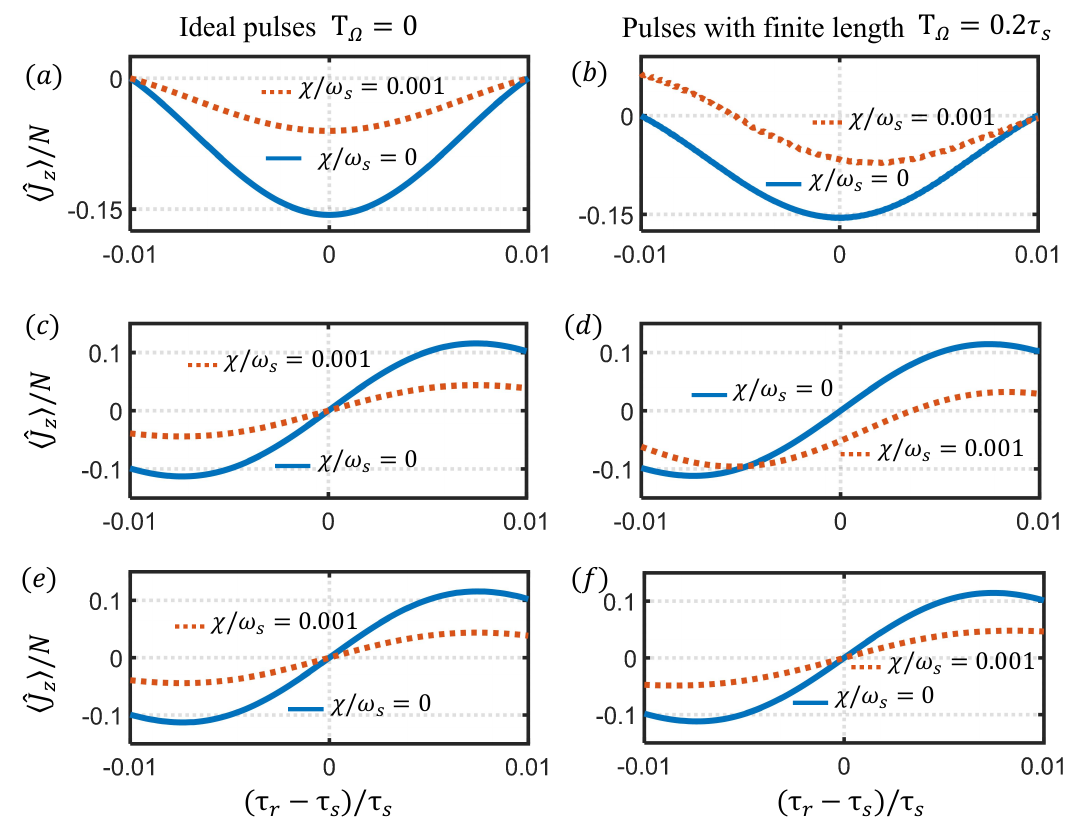}
\caption{\label{Fig3}(color online).
Spectra of many-body lock-in amplifiers with ideal (left column - $T_{\Omega}=0$) and finite-length (right column - $T_{\Omega}=0.2\tau_s$) pulses for $\chi=0$ (blue solid line) and $\chi=0.2\pi$ (red dotted line).
(a) and (b): using PDD sequence as reference signal.
(c) and (d): using CP sequence as reference signal.
(e) and (f): many-body lock-in amplifier using CP sequence along $y$ direction as reference signal via SPDMBI.
Here, $\gamma_g=1$, $B_\textrm{ac}=1$, $\omega_s=200\pi$, $T_{\Omega}=0.2\tau_s$, $\Omega=4\pi$, $N=20$, pulse number $L=99$ for PDD and $L=100$ for CP.
}
\end{figure}

In the limit of $T_\Omega \to 0$, the $\pi$ pulses are ideal and are described by $\Omega_\pi(t)=\pi\sum_{l=1}^{L}\delta_D(t-(l-\lambda)\tau_r)$, where $\delta_D(t)$ is the Dirac function.
One can obtain $\sin[\alpha(t)]=0$ and $\cos[\alpha(t,\lambda)]
=\left\{
\begin{array}{rcl}
1  &&{2k\tau_\textrm{r}\leq t+\lambda\tau_\textrm{r} \leq(2k+1)\tau_\textrm{r}},\\
-1  &&{(2k+1)\tau_\textrm{r}\leq t+\lambda\tau_\textrm{r} \leq 2(k+1)\tau_\textrm{r}
	}
\end{array} \right.$
which is a square wave function.
%
According to Fourier series expansion $\cos[\alpha(t,\lambda)]=\sum_{k=1,\textrm{odd}}\frac{4}{k\pi}\sin(k\omega_rt+k\lambda\pi)$, one can simplify the Hamiltonian Eq.\eqref{HIlock0} as
\begin{eqnarray}\label{HLeff0}
\hat{H}_{L'}(\lambda)\approx\chi\hat{J}_z^2+\frac{2\gamma_g B_{ac}}{\pi}\sin[(\omega_s-\omega_r)t+\lambda\pi]
\end{eqnarray}
ignoring the high frequency terms with $\omega_r=\pi/\tau_r$~\cite{SCCP2024,MZQF2024}.
Therefore, CP sequence with $\lambda=1/2$, we have
\begin{eqnarray}\label{HLCP}
\hat{H}_{L'}^\textrm{eff,CP}&=&\frac{1}{L\tau_r}\int_{0}^{L\tau_r}\hat{H}_{L'}(\lambda=1/2)dt\\\nonumber
&\approx&\chi\hat{J}_z^2+\frac{2\gamma_g B_\textrm{ac}}{L\pi}\frac{\sin^2(L\omega_s\delta_\tau/2)}{\sin(\omega_s\delta_\tau/2)}\hat{J}_z
\end{eqnarray}
with $\delta_\tau=\tau_r-\tau_s$ and $\tau=\pi/\omega$.
Similarly, for PDD sequence with $\lambda=0$, we have
\begin{eqnarray}\label{HLPDD}
\hat{H}_{L'}^\textrm{eff,PDD}&=&\frac{1}{L\tau_r}\int_{0}^{L\tau_r}\hat{H}_{L'}(\lambda=0)dt\\\nonumber
&\approx&\chi\hat{J}_z^2+\frac{2\gamma_g B_\textrm{ac}}{L\pi}\frac{\sin(L\omega_s\delta_\tau)}{\omega_s\delta_\tau}\hat{J}_z.
\end{eqnarray}
Inputting the symmetric state $\ket{\Psi(0)}=\ket{\Psi}_S$, the readout can be implemented with a $\frac{\pi}{2}$-pulse along $x$ direction: $\hat{U}_{r}=e^{-i\frac{\pi}{2}\hat{J}_{x}}$.
%
%
%
For the CP sequence with $\lambda=1/2$, the effective Hamiltonian Eq.~\eqref{HLCP} and the readout operator both satisfy Eq.~\eqref{H_sym} ensuring SPDMBI, so the lock-in signal $\langle J_z(t_L=L\tau_r)\rangle_{\delta_\tau}$ is antisymmetric about $\delta_\tau=\tau_r-\tau_s$ ($\tau_s=\pi/\omega_s$) [see Fig.~\ref{Fig3}~(c)].
While for the PDD sequence with $\lambda=0$, the effective Hamiltonian Eq.~\eqref{HLPDD} does not satisfy Eq.~\eqref{H_sym} and the corresponding spectrum is not a SPDMBI [see Fig.~\ref{Fig3}~(a)].

In practice, for pulses of the CP sequence with finite length $T_{\Omega}$, $\cos[\alpha(t)]=\sum_{k=1}^{+\infty}a_k\cos(k\omega_r t)$ and $\sin[\alpha(t)]=\sum_{k=1}^{+\infty}b_k\sin(k\omega_r t)$ with $a_k=\frac{2[1-(-1)^k]}{k\pi}\left[\sin\left(\frac{k\pi}{2}-\frac{k\pi T_{\Omega}}{2\tau_r}\right)+\frac{\cos\left(\frac{k+1}{2}\pi+\frac{k\pi T_{\Omega}}{2\tau_r}\right)}{1-\left(\frac{\tau_r}{kT_{\Omega}}\right)^2}\right]$ and $b_k=kT_{\Omega}a_k/\tau_r$.
Therefore the effective Hamiltonian for Eq.~\eqref{HIlock} becomes
\begin{eqnarray}\label{Heff10}
\small \hat{H}_\textrm{LA}^{\textrm{eff1}} &=& \frac{\chi}{2}(a_s\hat{J}_z^2 + b_s\hat{J}_y^2)  +\frac{a_1 \gamma_g B_\textrm{ac}}{2} \frac{\sin^2(L\omega_s\delta_\tau/2)}{L\omega_s\delta_\tau/2} \hat{J}_z\nonumber \\
&& + \frac{b_1 \gamma_g B_\textrm{ac}}{2} \frac{\sin(L\omega_s\delta_\tau)}{L\omega_s\delta_\tau}\hat{J}_y,
\end{eqnarray}
with $a_s=\sum_ka_k^2$, $b_s=\sum_kb_k^2$, $a_1=\frac{4\left(\frac{\tau_r}{T_{\Omega}}\right)^2\cos\left(\frac{\pi T_\Omega}{2\tau_r}\right)}{\pi\left[\left(\frac{\tau_r}{T_\Omega}\right)^2-1\right]}$ and $b_1=\frac{4\left(\frac{\tau_r}{T_{\Omega}}\right)\cos\left(\frac{\pi T_{\Omega}}{2\tau_r}\right)}{\pi\left[\left(\frac{\tau_r}{T_{\Omega}}\right)^2-1\right]}$.
Obviously, the last term in Eq.~\eqref{Heff10} breaks the symmetry in Eq.~\eqref{H_sym}.
Thus, when $\chi\neq0$, it will alter the shape and amplify the shift of the spectra, leading to an inaccurate estimation of the target frequency $\omega_s$, see Fig.~\ref{Fig3}~(d). 
Similar phenomena also occur in the case of PDD sequence, see Fig.~\ref{Fig3}~(b).

To accurately estimate $\omega_s$, we choose a series of equally spaced $\pi$ pulses in the $y$ direction as a reference signal, that is, the reference Hamiltonian is $\hat{H}_\textrm{ref}=\Omega_\pi(t)\hat{J}_y$.
Similarly, in the interaction picture with respect to $\hat{H}_\textrm{ref}=\Omega_\pi(t)\hat{J}_{y}$, the original Hamiltonian $\hat{H}_\textrm{LY}^0(t)=\chi\hat{J}_z^2+M(t)\hat{J}_z+\Omega_\pi(t)\hat J_y$ transforms into
\begin{equation}\label{HIlock}
 \hat{H}_\textrm{LY}(t)=\chi\hat{J}_\alpha'^2(t)+M(t)\hat{J}_\alpha'(t),
\end{equation}
where $\hat{J}_\alpha(t)=\cos[\alpha(t)]\hat{J}_z-\sin[\alpha(t)]\hat{J}_x$ and $\alpha(t)=\int_0^t\Omega_{\pi}(t')dt'$.
Thus the effective Hamiltonian for Eq.~\eqref{HIlock} becomes 
\begin{eqnarray}\label{Heff11}
\small \hat{H}_\textrm{LY}^{\textrm{eff1}} &=& \frac{\chi}{2}(a_s\hat{J}_z^2 + b_s\hat{J}_x^2)  +\frac{a_1 \gamma_g B_\textrm{ac}}{2} \frac{\sin^2(L\omega_s\delta_\tau/2)}{L\omega_s\delta_\tau/2} \hat{J}_z\nonumber \\
&& - \frac{b_1 \gamma_g B_\textrm{ac}}{2} \frac{\sin(L\omega_s\delta_\tau)}{L\omega_s\delta_\tau}\hat{J}_x
\end{eqnarray}
which satisfies Eq.~\eqref{H_sym}.
%
Thus the lock-in signal $\langle J_z(t_L)\rangle_{\delta_\tau}$ becomes antisymmetric about $\delta_\tau$ even taking pulse length into account for all $\chi$, see Fig.~\ref{Fig3}~(e) and (f).
%
%
%
%

{\it Conclusions and discussions. --}
In summary, we introduce the SPDMBI-based Ramsey spectroscopy that improves the robustness against inter-particle interactions and most experimental imperfections, such as, timing inaccuracies, decoherence and pulse length.
We demonstrate several key applications, including the measurement of time-independent signals and time-dependent signals using many-body lock-in amplifiers.
Notably, our protocol can be implemented without stringent timing control~\cite{JHarxiv2022} and is applicable to both non-interacting and interacting systems, encompassing Bose condensed atoms~\cite{ADCRMP2009,MWMRMP2020}, trapped ions~\cite{DLRMP2003,CMRMP2021}, nitrogen-vacancy centers in diamond~\cite{JFBRMP2020,JDRMP2024}, doped spins in semiconductors~\cite{GBRMP2023}, and even ``artificial atoms'' like superconducting qubits~\cite{ABRMP2021,JBNP2011,YSPRAppl2019,NEPRAppl2023}. 
Furthermore, employing symmetric entangled states such as spin-squeezed states, spin cat states and Greenberger-Horne-Zeilinger (GHZ) states, measurement precision can be further improved~\cite{SCPRA2025}.
Moreover, our SPDMBI protocol is general and can also be applied in Rabi spectroscopy~\cite{JHarxiv2022}.
Our protocol thus offers a pathway to enhance the performance of various quantum sensors, including magnetometers~\cite{CLDRMP2017,JFBRMP2020,MWMRMP2020,YSGRMP2000}, atomic clocks~\cite{ADLRMP2015,LPRMP2018}, weak-force detectors~\cite{RSNC2017}, and noise spectroscopy detectors~\cite{JBNP2011,IAPRA2016,SDCP2020,CJPRA2025}.
\acknowledgments{Sijie Chen and Jiahao Huang contribute equally. This work is supported by the National Natural Science Foundation of China (Grants No. 12025509, No12305022, and No. 92476201), the National Key Research and Development Program of China (Grant No. 2022YFA1404104), and the Guangdong Provincial Quantum Science Strate
gic Initiative (GDZX2305006 and GDZX2405002).}

\end{document}